\documentclass[10pt,a4paper]{article}
\date{}
\usepackage[english]{babel}
\usepackage{amssymb}
\title{Lagrangian Approach to O(2N) Model\\\ with $\theta$-deformed Target Space}
\author{$\textrm{A. Jahan}^1$, $\textrm{M. Nasseri}^{2}$\\
Research Institute for Astronomy and Astrophysics of Maragha, P. O. Box: 55134 - 44, IRAN\\\ Islamic Azad University-Hashtgerd branch, Tehran, $\textrm{Iran}^2$\\\
jahan@riaam.ac.ir}
\begin{document}
\maketitle
\begin{abstract}
We use the lagrangian approach to derive the free energy of a O(2N) model with $\theta$-deformed target space at finite temperature.
\end{abstract}
\section{Introduction}
Most of the recent studies in noncommutative (NC) field theories [1, 2, 3, 4] consider the NC space-time and assume the field space of the underlying theories to be commutative. As a result the noncommutativity reveals  itself through the interaction vertices and thus via the interactions. So for a  free non-interacting
theories there are no modifications arising from the noncommutativity of space-time. However, the situation changes when the noncommutativity is
implemented in target space manifold rather than the space-time manifold. Therefore, in contrast to the case of NC space-time manifold
the free part of the action modifies when the NC target space manifold is considered. The other characteristics of the theories with NC target space is their violation from the Lorentz invariance [5, 6]. In [7] authors have studied the black body spectrum of a massless sigma model with NC target manifold. Generalization to the U(1) gauge field is considered in [8]. The most recent studies involve the perturbative aspects and the Casimir effect in such models [9, 10].\\
In this letter generalize our previous work [11] to consider a O(2N) model with $\theta$-deformed field space in 1+1 dimensions
and shall apply the path integral approach to the NC quantum mechanics developed earlier [12, 13, 14, 15] to derive the finite temperature partition function of the model. Our result is in accordance with the expression derived for
the partition function by applying the method based on the Hamiltonian formalism [7]. As is expected, in the $\theta\rightarrow 0$ limit, we recover the partition function of the
model with usual commutative field space.
\section{O(2N) Model in Two Dimensions}
We consider a two dimensional O(2N) model with action
\begin{equation}\label{1}
\mathcal{S}=g\int{d^{2}x}\,\partial_{\alpha}\phi^{A}\partial^{\alpha}\phi^{A},
\end{equation}
with $\alpha=1,2$ and $A=1,\ldots,2N$. The the corresponding Lagrangian reads
\begin{equation}\label{2}
\mathcal{L}=g\int{dx}\Big(\dot{\phi}^{A}\dot{\phi}^{A}-\phi^{A}_{x}\phi^{A}_{x}\Big),\quad\quad\quad\phi^{A}_{x}\equiv\partial_{x}\phi^{A}.
\end{equation}
The field $\phi^{A}$ and it's conjugate momenta $\pi^{A}$ fulfils the equal-time canonical structure (setting $\hbar=1$)
{\setlength\arraycolsep{2pt}
\begin{eqnarray}\label{3}
[\phi^{A}(t,x),\phi^{A}(t,x^{\prime})]&=&i\delta^{AB}\delta(x-x^{\prime}),\\\nonumber
[\phi^{A}(t,x),\phi^{B}(t,x^{\prime})]&=&0,\\\nonumber
[\pi^{A}(t,x),\pi^{B}(t,x^{\prime})]&=&0.
\end{eqnarray}}
For the fields defined over the length $l(=1)$, we can expand them in terms of the normal modes $\chi_{n}^{A}(t)$ as
\begin{equation}\label{4}
\phi^{A}(t,x)=\sum^{\infty}_{n=1}\chi_{n}^{A}(t)\sin(\pi{}nx),
\end{equation}
with immediate result fore the Lagrangian and Hamiltonian as
\begin{equation}\label{5}
L=\frac{g}{2}\sum_{n}\Big(\dot{\chi}^{A}_{n}\dot{\chi}^{A}_{n}-\omega^{2}
_{n}\chi^{A}_{n}\chi^{A}_{n}\Big),
\end{equation}
and
\begin{equation}\label{6}
H=\frac{1}{2g}\sum_{n}\Big(p^{A}_{n}p^{A}_{n}+g^{2}\omega^{2}
_{n}\chi^{A}_{n}\chi^{A}_{n}\Big).
\end{equation}
where $\omega_{n}=\pi{n}$. Now the expressions (5) and (6) can be interpreted as the Lagrangian and Hamiltonian of an infinite series of uncoupled
harmonic oscillators with mass \textit{g}.
\section{$\theta$-deformed Target Manifold}
$\theta$-deformation of the target manifold is implemented via the deformation of canonical structure to a noncommutative one
{\setlength\arraycolsep{2pt}
\begin{eqnarray}\label{7}
[\hat{\phi}^{A}(t,x),\hat{\pi}^{B}(t,x^{\prime})]&=&i\delta^{AB}\delta(x-x^{\prime}),\\\nonumber
[\hat{\phi}^{I_k}(t,x),\hat{\phi}^{I^\prime_k}(t,x^{\prime})]&=&i\theta\varepsilon^{I_kI^\prime_k}\delta(x-x^{\prime}),\\\nonumber
[\hat{\pi}^{A}(t,x),\hat{\pi}^{B}(t,x^{\prime})]&=&0,
\end{eqnarray}}
For the antisymmetric matrix $\varepsilon^{I_kI^\prime_k}$ we assume $\varepsilon^{I_kI^\prime_k}=1$ if $I_k<I^\prime_k$. Here $I_k,I^\prime_k\in\{2k-1,2k\}$ and thus $A,B\in\bigcup_{k=1}^{N}I_k$. Then from Eq.(7) one finds the corresponding structure for the expansion modes as
{\setlength\arraycolsep{2pt}
\begin{eqnarray}\label{8}
[\hat{\chi}^{A}_{n}(t),\hat{p}^{B}_{m}(t)]&=&i\delta^{AB}\delta_{n,m},\\\nonumber
[\hat{\chi}^{I_k}_{n}(t),\hat{\chi}^{I^\prime_k}_{m}(t)]&=&i\theta\varepsilon^{I_kI^\prime_k}\delta_{n,m},\\\nonumber
[\hat{p}^{A}_{n}(t),\hat{p}^{B}_{m}(t)]&=&0.
\end{eqnarray}}
Note that we have considered a block-diagonal form for the noncommutativty matrix $\theta^{AB}$. This implies that the noncommutativty is assumed just for two adjacent coordinates, i.e.
{\setlength\arraycolsep{2pt}
\begin{eqnarray}\label{9}
[\phi^{2k-1}(t,x),\phi^{2k}(t,x^{\prime})]=\cdots=i\theta\delta(x-x^{\prime}),\qquad k=1,\ldots,N
\end{eqnarray}}
However, the usual canonical structure (3) is recovered, if we redefine the noncommutative field $\hat{\phi}^{A}(x,t)$ in terms of the commutative fields $\phi^{A}(x,t)$ via
{\setlength\arraycolsep{2pt}
\begin{eqnarray}\label{10}
\hat{\phi}^{I_k}(x,t)&=&\phi^{I_k}(x,t)-\frac{\theta}{2}\epsilon^{I_kI^\prime_k}\pi^{I^\prime_k}(x,t),\\
\hat{\pi}^{A}(x,t)&=&\pi^{A}(x,t),\nonumber
\end{eqnarray}}
and
{\setlength\arraycolsep{2pt}
\begin{eqnarray}\label{10}
\hat{\chi}^{I_k}_{n}(t)&=&\chi^{I_k}_{n}(t)-\frac{\theta}{2}\epsilon^{I_kI^\prime_k}p^{I^\prime_k}_{n}(t),\\
\hat{p}^{A}_{n}(t)&=&p^{A}_{n}(t).\nonumber
\end{eqnarray}}
Therefore, upon substituting for the noncommutative fields from (10) in (5) and (6) we find the $\theta$-deformed Hamiltonian and Lagrangian as [9, 10, 11]
\begin{equation}\label{12}
H_{\theta}=\sum_{n_k}\bigg(\frac{\kappa_{n_k}}{2g}p^{I_k}_{n_k}p^{I_k}_{n_k}+\frac{g}{2}\omega^{2}_
{n_k}\chi^{I_k}_{n_k}\chi^{I_k}_{n_k}-\frac{g}{2}\theta\omega^{2}_{n_k}\epsilon^{I_kI_k^\prime}
\chi^{I_k}_{n_k}p^{I^\prime_k}_{n_k}\bigg),
\end{equation}
and
\begin{equation}\label{13}
L_{\theta}=\sum_{n_k}\bigg(\frac{g}{2\kappa_{n_k}}\dot{\chi}^{I_k}_{n_k}\dot{\chi}^{I_k}
_{n_k}-\frac{g\omega^{2}_{n_k}}{2\kappa_{n_k}}\chi^{I_k}_{n_k}\chi^{I_k}_{n_k}
+\frac{g^{2}\omega^{2}_{n_k}\theta}{2\kappa_{n_k}}\epsilon^{I_kI_k^\prime}
\chi^{I_k}_{n_k}\dot{\chi}^{I_k^\prime}_{n_k}\bigg),
\end{equation}
with the $\theta$-deformed action as
{\setlength\arraycolsep{2pt}
\begin{eqnarray}\label{14}
{S}_{\theta}=\sum_{I_k}\sum_{n_k}\tilde{{S}}_{\theta}[\chi_{n_k}^{I_k}]=\frac{g}{2}\sum_{n_k}\int{dt}\bigg(\frac{1}{\kappa_{n_k}}\dot{\chi}^{I_k}_{n_k}\dot{\chi}^{I_k}
_{n_k}-\frac{\omega^{2}_{n_k}}{\kappa_{n_k}}\chi^{I_k}_{n_k}\chi^{I_k}_{n_k}
+\frac{g\theta\omega^{2}_{n_k}}{\kappa_{n_k}}\epsilon^{I_kI_k^\prime}
\chi^{I_k}_{n_k}\dot{\chi}^{I^\prime_k}_{n_k}\bigg)
\end{eqnarray}}
where $\kappa_n=1+\frac{1}{4}g^2\theta^2\omega^2_n$.
\section{Finite Temperature Partition Function}
For the Euclidean-time action the finite temperature partition function is [9, 10, 11, 12]
{\setlength\arraycolsep{2pt}
\begin{eqnarray}\label{15}
\mathcal Z_{\theta}(\beta)&=&\prod_{I_k}\Bigg(\int \prod_{n_k}d\chi^{I_k}_{n_k}e^{-{S}_{\theta}}\Bigg)=\Bigg(\int \prod_{n}d\chi^{I_1}_{n}e^{-\sum_n\tilde{S}_{\theta}[\chi^{I_1}_{n}]}\Bigg)^N\\\nonumber
&=&\Big(\prod_{n}TrK_{\theta}(\vec{\chi}_{n}^{\,\prime\prime},\beta;\vec{\chi}_{n}^{\,\prime},0)\Big)^N
\end{eqnarray}}
where the single-particle propagator is given by
\begin{equation}\label{16}
K_{\theta}(\vec{\chi}_{n}^{\,\prime\prime},\beta;\vec{\chi}_{n}^{\,\prime},0)=
\frac{g\omega_{n}}{2\pi\sqrt{\kappa_{n}}\sinh(\beta\omega_{n}\sqrt{\kappa_{n}})}
e^{-A(\vec{\chi}^{\,\prime\prime}_{n},\beta;\vec{\chi}^{\,\prime}_{n},0)}.\nonumber
\end{equation}
with $\vec{\chi}_{n}=(\chi^{1}_{n},\chi^{2}_{n})$ and
{\setlength\arraycolsep{2pt}
\begin{eqnarray}\label{17}
A(\vec{\chi}^{\,\prime\prime}_{n},\beta;\vec{\chi}^{\,\prime}_{n},0)&=&\frac{g}{2}
\int^{\beta}_{0}{d\tau}\bigg(\frac{1}{\kappa_{n}}\dot{\chi}^{a}_
{n}\dot{\chi}^{a}
_{n}+\frac{\omega^{2}_{n}}{\kappa_{n}}\chi^{a}_{n}\chi^{a}_{n}
+\frac{g\omega^{2}_{n}}{i\kappa_{n}}\theta\epsilon^{ab}
\chi^{a}_{n}\dot{\chi}^{b}_{n}\bigg),\\\nonumber
&=&\frac{g\omega_{n}}{2\sqrt{\kappa_{n}}\sinh(\beta\omega_{n}\sqrt\kappa_{n})}
\bigg[(\vec{\chi}^{\,\prime\prime\,2}_{n}+
\vec{\chi}^{\,\prime\,2}_{n})\cosh(\beta\omega_{n}\sqrt{\kappa_{n}})\\
&-&2(\vec{\chi}^{\,\prime}_{n}\cdot{\vec{\chi}^{\,\prime\prime}_{n}})
\cosh(\beta\omega_{n}\sqrt{\kappa_{n}-1})\nonumber
+2(\vec{\chi}^{\,\prime}_{n}\times{\vec{\chi}^{\,\prime\prime}_{n}})_{z}
\sinh(\beta\omega_{n}\sqrt{\kappa_{n}-1})\bigg],\nonumber
\end{eqnarray}}
where $\vec{\chi}^{\,\prime\prime}_{n}=\vec{\chi}_{n}(\beta)$ and $\vec{\chi}^{\,\prime}_{n}=\vec{\chi}_{n}(0)$. Also we have assumed $a,b=1,2$. Here the dots stand for the derivative with respect to $\tau(=it)$.
For the trace of the exponential term in (16) we find
\begin{equation}\label{18}
Tre^{-A}=\frac{\pi\sqrt{\kappa}_{n}}{g\omega_{n}}
\frac{\sinh(\beta\omega_{n}\sqrt{\kappa_{n}})}{\cosh(\beta\omega_{n}\sqrt{\kappa_{n}})
-\cosh(\beta\omega_{n}\sqrt{\kappa_{n}-1})}.
\end{equation}
where the functional trace is defined to be
\begin{equation}\label{19}
Trf(x,y)=\int dx f(x,x)
\end{equation}
Therefore one is left with the partition function as
{\setlength\arraycolsep{2pt}
\begin{eqnarray}\label{20}
\mathcal{Z}_{\theta}(\beta)&=&\Bigg(\prod_{n}\frac{g\omega_{n}}{2\pi\sqrt{\kappa_{n}}\sinh(\beta\omega_{n}\sqrt{\kappa}_{n})}
Tre^{-A}\Bigg)^{N},\\
&=&\Bigg(\prod_{n}\frac{1}{4\sinh\big(\beta\omega_{n}\gamma_{+,n}\big)
\sinh\big(\beta\omega_{n}\gamma_{-,n}\big)}\Bigg)^{N}.\nonumber
\end{eqnarray}}
where $\gamma_{\pm,n}=\frac{1}{2}(\sqrt{\kappa_{n}}\pm\sqrt{\kappa_{n}-1}\,)$. Hence by excluding the divergent contribution $N\sum_n\omega_n\sqrt{\kappa_n}$ one finds the free energy $\mathcal{F}_{\theta}(\beta)$ as
{\setlength\arraycolsep{2pt}
\begin{eqnarray}\label{21}
\mathcal{F}_{\theta}(\beta)&=&-\frac{1}{\beta}\textmd{ln}\mathcal {Z}_{\theta}(\beta),\\
&=&N\sum_{n}\Big[\textmd{ln}(1-e^{-2\beta\omega_{n}\gamma_{-,n}})+
\textmd{ln}(1-e^{-2\beta\omega_{n}\gamma_{+,n}})\Big].\nonumber
\end{eqnarray}}
Now, with $N=1$ the above result coincides with one which was derived earlier in [7] by employing the Hamiltonian formalism. In $\theta\rightarrow{0}$ limit, $\gamma_{\pm,n}$ tends to $\frac{1}{2}$ and one recover the usual free energy as
\begin{equation}\label{22}
\mathcal{F}(\beta)=2N\sum_{n}\textmd{ln}\big(1-e^{-\beta\omega_{n}}\big).
\end{equation}
The $\theta$-deformed partition function up to first order in noncommutativity parameter reads
\begin{equation}\label{23}
\mathcal Z_\beta=e^{\frac{\beta\pi}{12}N}\prod_n\Big(1-e^{-\beta\Omega_n+}\Big)^{N}\Big(1-e^{-\beta\Omega_n-}\Big)^{N}
\end{equation}
where we have invoked the Riemann zeta function $\zeta(-1)=\frac{1}{12}$ and $\Omega_{n\pm}=\omega_n(1\pm\frac{1}{2}g\theta\omega_n)$.
\section{Acknowledgments}
Authors would like to thank Islamic Azad University of Hashtgerd for their support.


\begin{thebibliography}{99}
\bibitem{1}R. Szabo,
\emph{Quantum field theory on
noncommutative spaces},
Phys. Rept. 378 (2003) 207-299.
\bibitem{2} M. Wohlgenannt,
\emph{Noncommutative Geometry and Physics},
arXiv: hep-th/0602105.
\bibitem{3}
M. Wohlgenannt,
\emph{Introduction to a noncommutative version of the standard model},
arXiv: hepth/0302070.
\bibitem{4}
H. O. Girotti,
\emph{Noncommutative Quantum Field Theory},
arXiv: hep-th/0301237.
\bibitem{5}
J. M. Carmona, J. L. Cortes, J. Gamboa, F. Mendez,
\emph{Quantum Theory of Noncommutative Fields},
JHEP 0303 (2003) 058.
\bibitem{6}
J. M. Carmona, J. L. Cortes, J. Gamboa, F. Mendez,
\emph{Noncommutativity in Field Space and Lorentz Invariance Violation}.
Phys. Lett B 565 (2003) 222-228.
\bibitem{7}
A. P. Balachandran, A. R. Queiroz, A. M. Marques, P. Teotonio-Sobrinho,
\emph{Quantum Fields with Noncommutative Target Spaces},
arXiv: hep-th/0706007v3.
\bibitem{8}
A. Fatollahi, M. Hajirahimi,
\emph{Black Body Radiation of Noncommutative Gauge Fields },
Phys. Lett. B 641 (2006).
\bibitem{10}
F. Khelili,
\emph{Noncommutative Complex Scalar Field and Casimir Effect},
Phys. Rev D85, 125013.
\bibitem{11}
F. Khelili,
\emph{Path Integral Quantization of Noncommutative Complex Scalar Field},
http://arxiv.org/abs/1109.4741.
\bibitem{11}
A. Jahan,
\emph{Path Iintegral Formulation of sigma model with noncommutative field space},
Fizika B18 (2009) 4, 189-194.
\bibitem{9} C. Acatrinei,
\emph{Path Integral Formulation of Noncommutative Quantum Mechanics},
JHEP 0109 (2001) 007.
\bibitem{12}
B. Dragovic, Z. Rakic,
\emph{Path Integrals in Noncommutative Quantum Mechanics},
Theor. Math. Phys. 140 (2004) 1299-1038.
\bibitem{13}
B. Dragovic, Z. Rakic,
\emph{Path Integral Approach to Noncommutative Quantum Mechanics},
arXiv: hep-th/0401198.
\bibitem{14}
A. Jahan,
\emph{Noncommutative Harmonic Oscillator at Finite Temperature: A Path Integral Approach},
Braz. J. Phys 38 (2008).
\end{thebibliography}
\end{document}